\documentclass[letterpaper, 10 pt, conference]{ieeeconf}  

\IEEEoverridecommandlockouts            

\usepackage{graphicx}

\title{\LARGE \bf
FlyHaptics: Flying Multi-contact Haptic Interface 
}

\author{Luis Moreno, Miguel Altamirano Cabrera, Muhammad Haris Khan,\\ Issatay Tokmurziyev, Yara Mahmoud,  Valerii Serpiva, and Dzmitry Tsetserukou\\
\thanks{The authors are with the Intelligent Space Robotics Laboratory, Center for Digital Engineering, Skolkovo Institute of Science and Technology. 
{\tt \{Luis.Moreno, m.altamirano, Haris.Khan, Issatay.Tokmurziyev, Valerii.Serpiva d.tsetserukou\}@skoltech.ru}}
}

\begin{document}

\maketitle
\thispagestyle{empty}
\pagestyle{empty}

\begin{abstract}

This work presents FlyHaptics, an aerial haptic interface tracked via a Vicon optical motion capture system and built around six five-bar linkage assemblies enclosed in a lightweight protective cage. We predefined five static tactile patterns—each characterized by distinct combinations of linkage contact points and vibration intensities—and evaluated them in a grounded pilot study, where participants achieved 86.5  recognition accuracy (F(4, 35) = 1.47, p = 0.23) with no significant differences between patterns. Complementary flight demonstrations confirmed stable hover performance and consistent force output under realistic operating conditions. These pilot results validate the feasibility of drone-mounted, multi-contact haptic feedback and lay the groundwork for future integration into fully immersive VR, teleoperation, and remote interaction scenarios.

\end{abstract}

\section{INTRODUCTION}

Haptic interfaces bridge the gap between physical and virtual environments by enabling users to experience tactile feedback. Traditional haptic systems, such as wearable and grounded devices, have been widely explored to enhance immersion in virtual reality (VR), teleoperation, and remote collaboration. Wearable systems offer flexibility in feedback delivery but can be bulky, restrict movement, or reduce feedback intensity due to size and weight constraints \cite{Head}. For example, finger-mounted or palm-based devices have been proposed to provide localized sensations, though they often compromise ergonomics or power \cite{Fingeret, HandPalm}. Grounded haptic displays, on the other hand, provide accurate and strong force feedback—often through robotic arms or exoskeletons—but limit interaction to a fixed workspace \cite{Teslamirror}. This restricts their applicability in untethered or large-area VR scenarios \cite{pocopo}.

To overcome these constraints, aerial haptics has emerged as a promising alternative. By using drones to deliver force feedback without requiring physical attachment, these systems offer greater freedom of movement and spatial interaction \cite{hapticbots}. However, most existing aerial haptic implementations are limited to single-point feedback delivered through rigid couplings or static configurations \cite{hapticdrone1point}, which restricts their ability to simulate rich, multi-contact tactile interactions. Some efforts, such as origami-inspired actuation, attempt to increase tactile variety through deformable structures, but they remain constrained by mechanical limits \cite{Origami}.

\begin{figure}[t]
\centerline{\includegraphics[width=0.4\textwidth, height=0.4\textwidth]{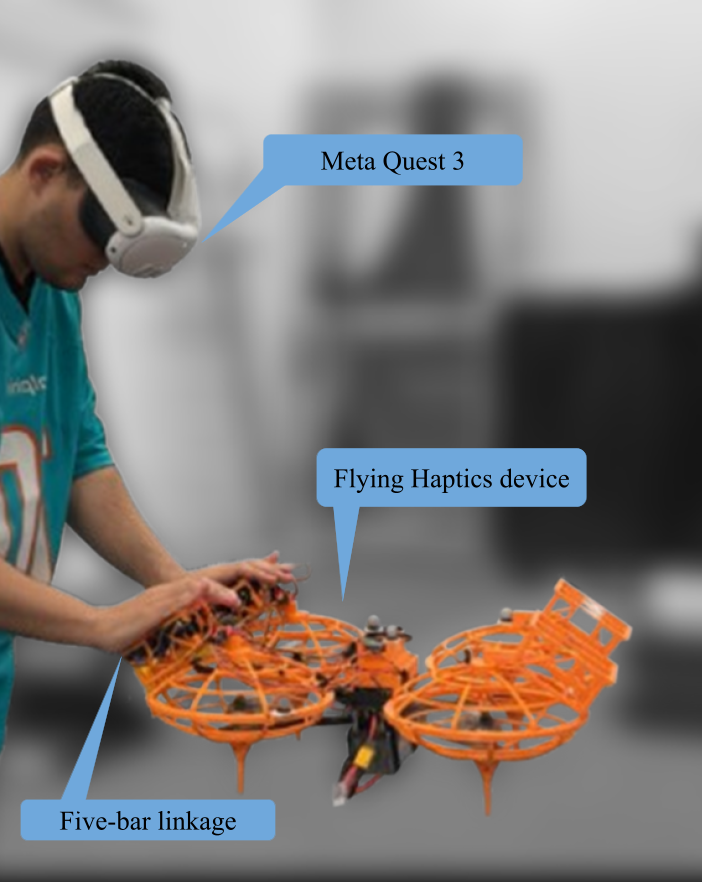}}
\caption{FlyHaptics delivers tactile feedback to VR user}
\label{fig:teaser}
\vspace{-0.4cm}
\end{figure}

We introduce FlyHaptics, a multi-contact drone-based haptic interface that integrates five-bar linkage mechanisms to generate diverse static and dynamic tactile patterns in mid-air. Unlike previous systems, FlyHaptics offers a novel design that enables aerial, multi-point tactile feedback, demonstrating the feasibility of delivering complex haptic interactions without physical constraints. This system lays the groundwork for future haptic applications that allow users to perceive surface textures, directional forces, and object contours—without relying on wearable devices. By combining precise motor control and an optimized mechanical structure, FlyHaptics has the potential to improve the realism of VR experiences and open new possibilities in teleoperation, remote collaboration, and augmented reality.

Although drones offer advantages such as mobility and unanchored operation, they present challenges in terms of flight stability, vibration suppression, and user safety \cite{dronehaptics}. FlyHaptics addresses these issues through the integration of a lightweight protective frame and tightly controlled actuation. This ensures that the system remains safe and stable while preserving its ability to deliver expressive, multi-contact tactile feedback \cite{GuideCopter}. We present FlyHaptics as a foundation for future integration into immersive VR environments, where intuitive and dynamic mid-air haptics could greatly enhance user experience.

\section{Related Work}

Haptic systems can be categorized into three primary branches: graspable, wearable\cite{tactile}, and touchable devices. Among these, touchable devices were the earliest to be developed and implemented, primarily as stationary systems designed to provide users with tactile feedback\cite{Musinger}. These systems include desktop-based screens that generate vibration-based responses, as well as stylus-like pens that deliver force feedback to simulate real-world interactions \cite{Pen}.

Touchable haptic devices have played a significant role in advancing telepresence technologies, enabling users to experience a sense of touch despite being physically distant from the object of interaction \cite{Teleop}. One of the most notable applications is in the medical domain—particularly in robotic-assisted surgery—where surgeons manipulate instruments remotely while receiving haptic feedback that replicates the sensation of contact with human tissue \cite{Surgery}. In industrial contexts, touchable haptic devices also support remote equipment maintenance, enabling operators to perform precision-based tasks with enhanced control and accuracy \cite{TouchableDisplay}.

While wearable and graspable haptic interfaces offer a wider range of interaction modalities, they remain inherently limited by their physical size, weight, and mechanical complexity \cite{WearableDevices}. To address these constraints, researchers have proposed encounter-type haptic displays, which allow users to receive haptic feedback without physically attaching to a device \cite{Encounter}. These systems expand the concept of haptic delivery by incorporating robotic manipulators—such as Universal Robots (UR) arms—or leveraging mobile robotic platforms and aerial vehicles as dynamic feedback tools \cite{CoboDeck}.

Encounter-type displays offer a key advantage: they eliminate constraints imposed by the physical form factor of the device itself \cite{SurveyEncounter}. For instance, robotic arms can deliver force feedback by repositioning themselves dynamically to simulate physical resistance, while mobile robots can render spatial cues along an x-y plane. However, drones provide a unique benefit over both—namely, the ability to deliver unconstrained 3D tactile feedback. Unlike robotic manipulators limited by reach, or ground robots confined to flat surfaces, drones can navigate freely through space, enabling more immersive and adaptive haptic interaction \cite{Manipulator}. This makes them especially promising for use in VR, telepresence, and remote operation scenarios \cite{Wired}.

Although drones have seen widespread adoption in industrial applications such as surveillance, inspection, logistics, and delivery, their use as haptic feedback systems remains relatively underexplored \cite{DronesIdustry}. Nonetheless, growing interest in this area has led researchers to explore innovative drone-based haptics for virtual and augmented reality environments \cite{HapticDrone}.

One notable approach involves using the drone’s own frame to provide direct force feedback, allowing users to physically interact with the device in mid-air \cite{HapticPuppet}. Another method mounts fixed haptic elements on the drone that serve as contact points for delivering tactile sensations during user interaction \cite{VRDrone}. Furthermore, swarm-based strategies have emerged, where multiple coordinated drones employ wire-tension systems to simulate object rendering and force feedback through distributed control \cite{Dandelion}. Collectively, these techniques highlight the evolving potential of drones as versatile, mid-air haptic interfaces—offering greater flexibility and freedom than traditional grounded or wearable alternatives.

\section{System Architecture}
FlyHaptics is a drone-based haptic interface designed to provide multi-contact force feedback in mid-air. The system integrates precise flight control, real-time localization, and haptic actuation, enabling users to experience realistic tactile sensations without the constraints of wearable or grounded devices. Built on the Robot Operating System (ROS) \cite{ROS}, the system runs on an Orange Pi 5B \cite{OrangePi}, which handles both haptic rendering and flight dynamics. For localization, a Vicon Tracking System is employed, offering high-precision tracking of both the drone and the user to maintain accurate positioning during operation.

\subsection{Overview of Drone for Haptic Device}

The FlyHaptics prototype (see Fig. \ref{fig:teaser}) is built on a custom 8-inch quadcopter designed for stable and responsive motion in haptic interaction scenarios. At the core of the drone is a SpeedyBee F405 flight controller, a lightweight unit responsible for processing flight commands and maintaining aerial stability. The controller operates using ArduPilot firmware \cite{ArduPilot}, an open-source platform widely adopted in autonomous and remotely piloted aircraft systems, providing robust support for dynamic control during mid-air tasks.

The quadcopter communicates with the high-level processor via MAVROS \cite{MAVROS}, a ROS-based bridge that facilitates message passing between the onboard flight controller and external computational units. In this configuration, the Orange Pi 5B computes desired trajectories and transmits position and velocity commands through MAVROS, enabling fine-grained motion control for responsive haptic delivery.

\begin{figure}[h]
\centerline{\includegraphics[width=0.5\textwidth]{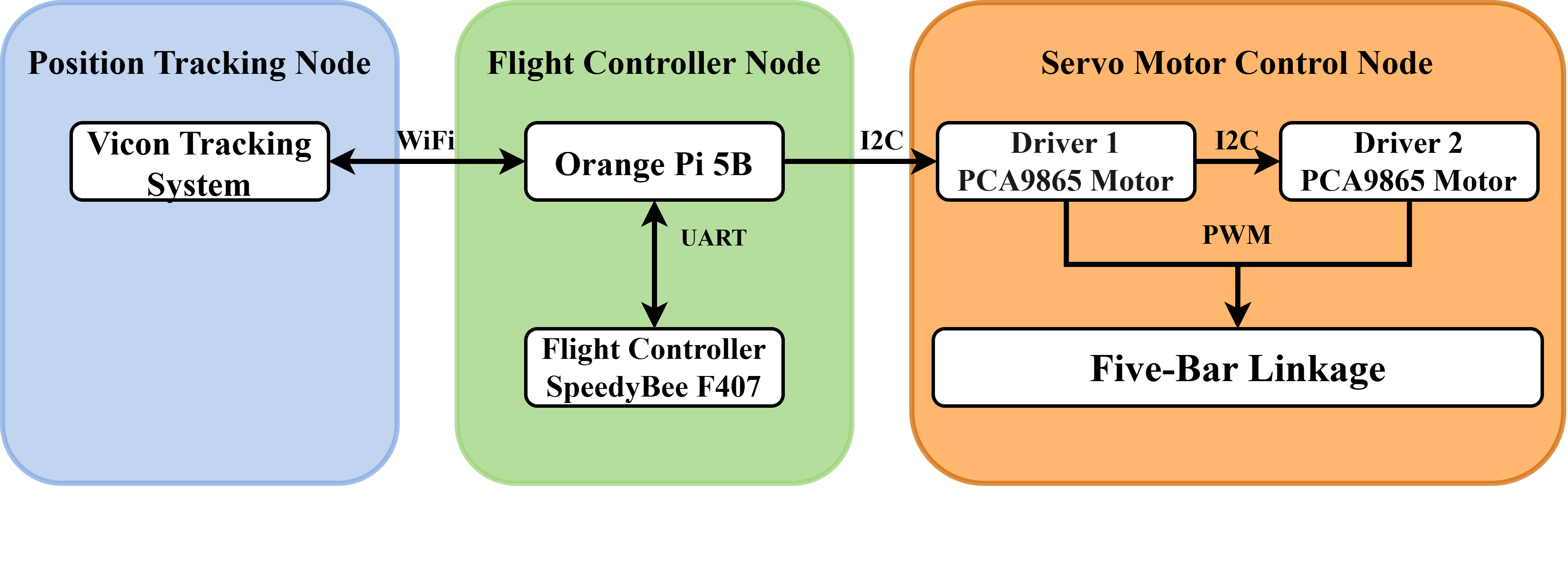}}
\caption{System Architecture}
\label{fig:System_Diagram}
\vspace{-0.4cm}
\end{figure}

To ensure stable and precise movement during user interaction, the system uses a Vicon motion capture setup composed of 14 infrared cameras. This configuration provides millimeter-level accuracy, essential for ensuring reliable positioning and minimal drift during haptic tasks. Real-time pose data from Vicon is continuously relayed to the onboard system to adjust flight behavior and maintain control fidelity. An overview of the complete system architecture is presented in Fig. \ref{fig:System_Diagram}.

\begin{figure}[t]
\centerline{\includegraphics[width=0.5\textwidth]{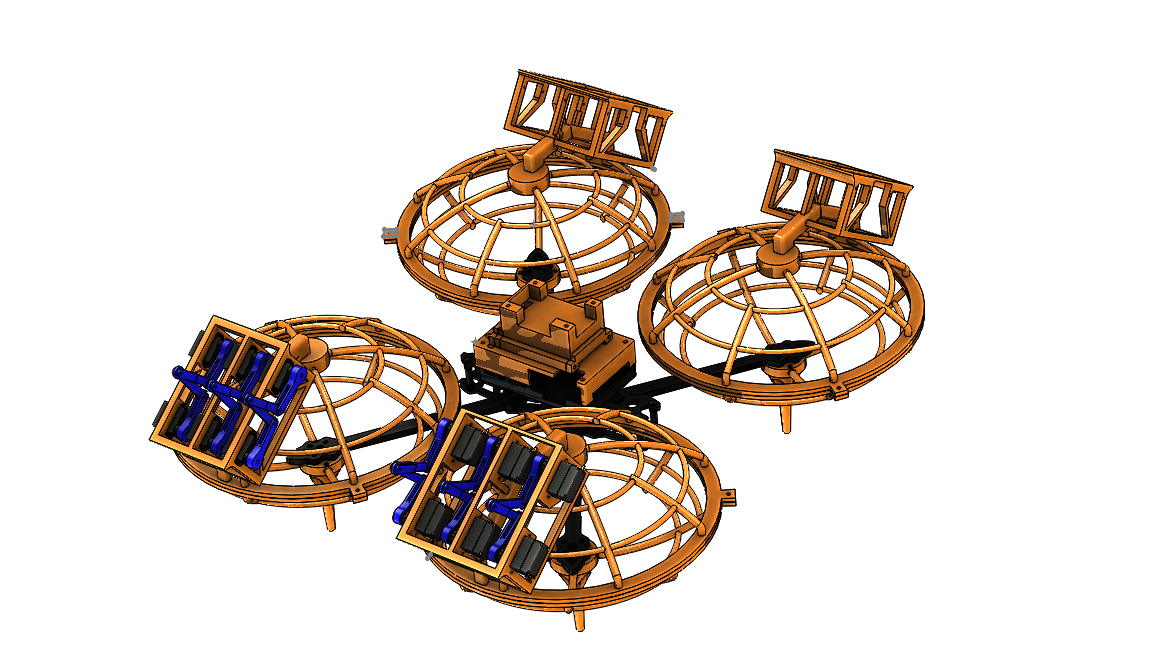}}
\caption{Haptic Drone CAD Design}
\label{fig:teaser}
\vspace{-0.4cm}
\end{figure}

\subsection{Integration of Haptic Device with Drone}

The FlyHaptics platform is built on a carbon fiber quadcopter frame, selected for its strength-to-weight ratio, providing mechanical stability without compromising agility. To safeguard users during interaction, each rotor is enclosed in a 3D-printed PLA protective cage. These enclosures serve to prevent accidental contact with the propellers while preserving aerodynamic performance.

Mounted on the top deck of the drone are servomotors and PWM drivers, responsible for actuating the five-bar linkage mechanisms. These mechanical assemblies generate force feedback by precisely adjusting their angles in response to control commands from the onboard processor. This modular integration allows FlyHaptics to deliver continuous tactile signals while maintaining flight balance.

\begin{figure}[t]
\centerline{\includegraphics[width=0.5\textwidth]{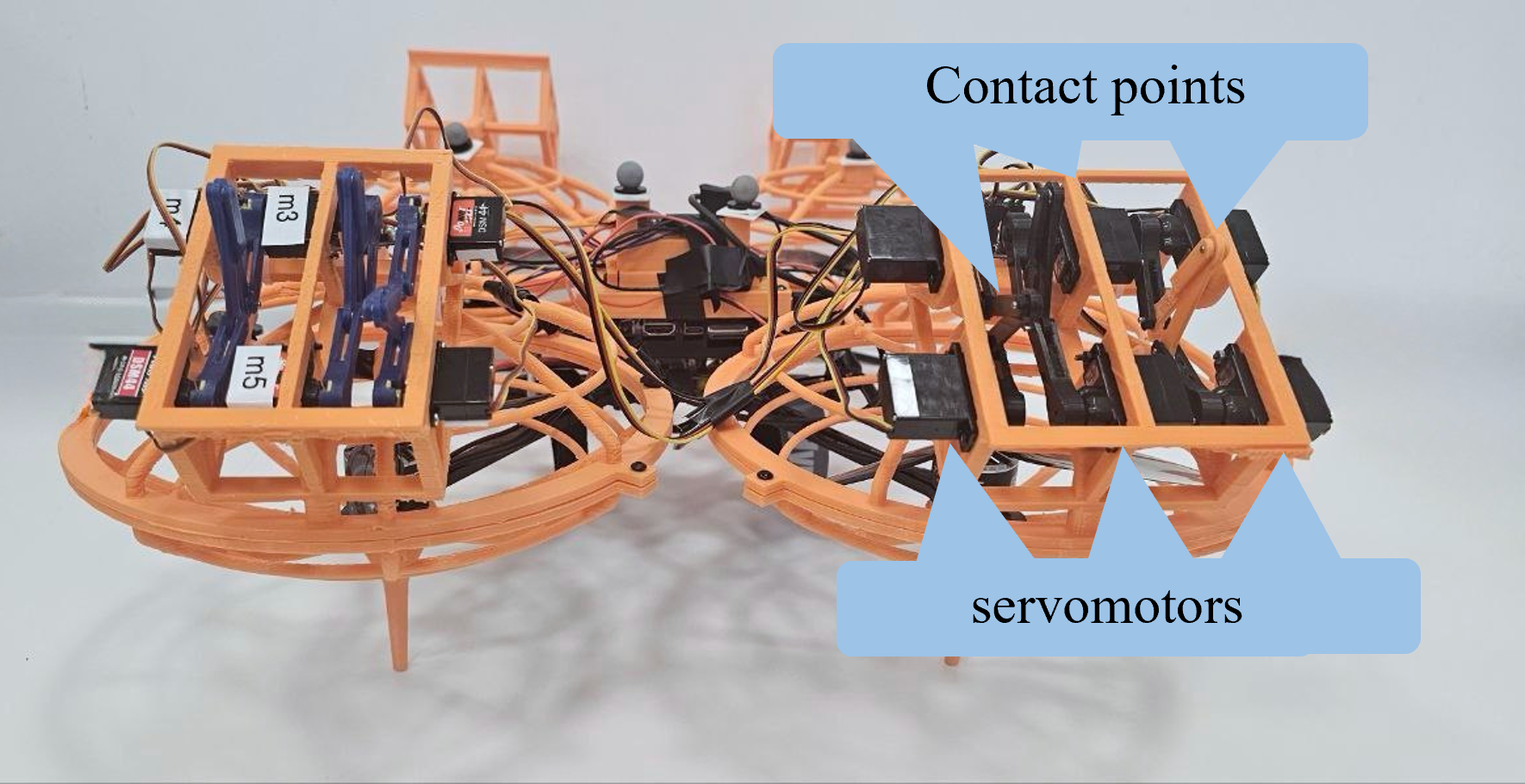}}
\caption{FlyHaptics Haptic Device}
\label{fig:haptics}
\vspace{-0.4cm}
\end{figure}

The system is powered by a six-cell Li-Po battery, which delivers raw power to the drone's flight controller while distributing regulated voltage to peripheral devices via a Universal Battery Elimination Circuit (UBEC). This power distribution ensures that components such as the Orange Pi 5B and the motor drivers receive stable current, preventing voltage drops that could impair control precision.

For communication between the controller and actuators, the system uses the I2C protocol, transmitting commands from the Orange Pi 5B to the PWM PCA9685 drivers. These drivers regulate signal timing to the DMS44 and HS-70MG servomotors, which were chosen for their compact size, torque reliability, and responsiveness. These actuators are critical to the movement of the five-bar linkage mechanisms, ensuring fast and accurate generation of haptic cues.

The haptic interface itself consists of six five-bar linkage assemblies. Each mechanism is designed to convert rotary motion into multi-directional end-effector displacement, enabling fine-tuned and expressive force feedback. By varying the position and orientation of these linkages, FlyHaptics can render different tactile sensations—such as surface texture variations, directional pressures, and resistance levels—resulting in a versatile haptic experience.

\subsection{Haptic Patterns}

\begin{figure}[h]
\centerline{\includegraphics[width=0.5\textwidth]{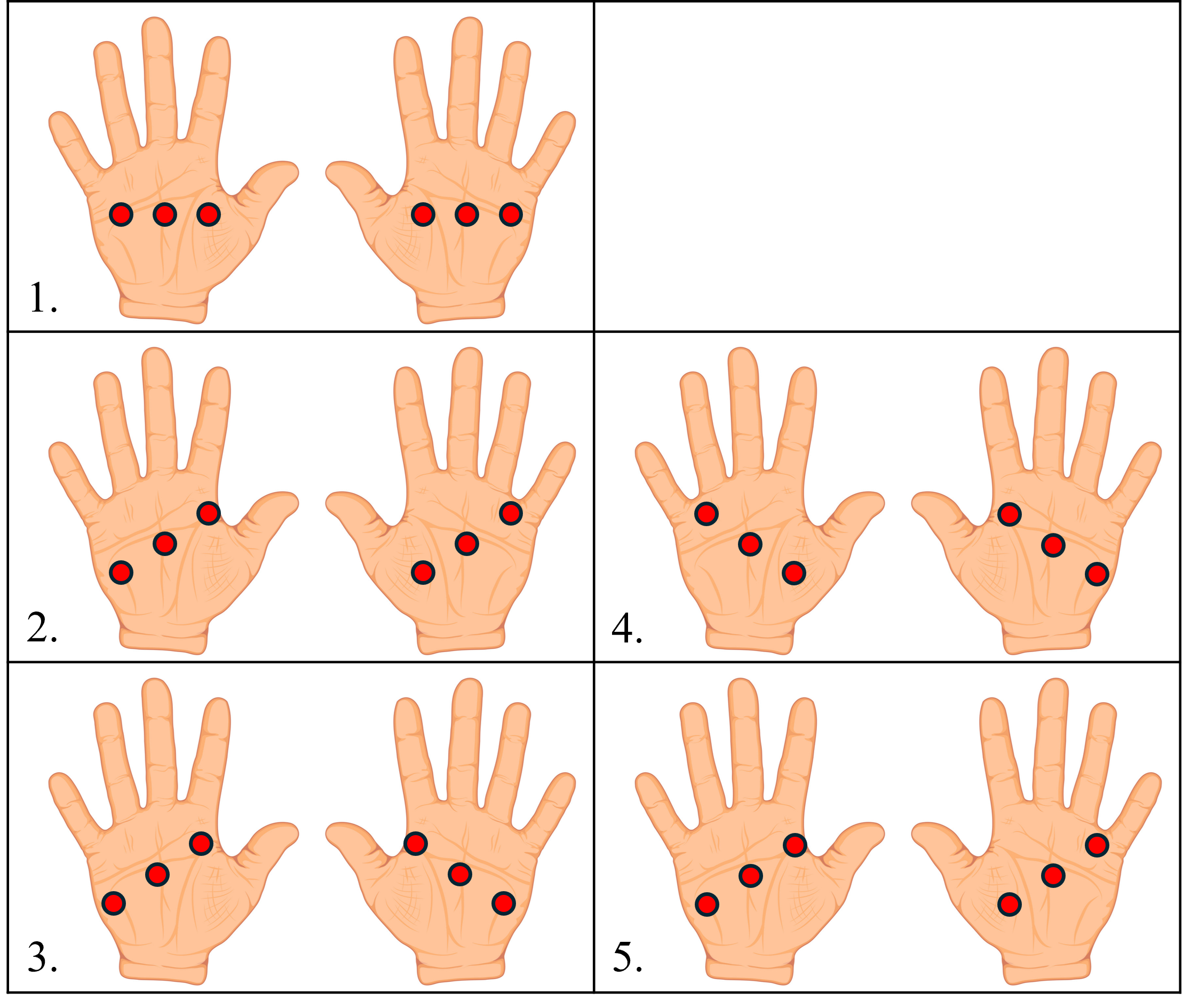}}
\caption{Haptic Patterns Diagram}
\label{fig:patterns}
\vspace{-0.4cm}
\end{figure}

The five-bar linkage mechanisms in FlyHaptics are capable of producing both static and dynamic haptic patterns. For the purposes of this study, the focus is on static patterns, which are applied simultaneously to both hands to ensure consistency during evaluation. Five distinct tactile patterns were designed, each intended to convey a unique spatial force profile without causing sensory overload. These patterns form the basis of the user study and are illustrated in Fig. \ref{fig:patterns}.

\section{User Study}

\subsection{Experimental Setup}

This study was conducted to evaluate participants' ability to perceive, distinguish, and accurately identify the predefined tactile patterns generated by the FlyHaptics system. The primary objective was to determine how effectively users could recognize different haptic cues, and whether variations in force and movement influenced their perception.

To eliminate potential confounding variables, all experiments were conducted in a controlled indoor setting. Environmental conditions such as background noise, lighting, and visual distractions were minimized, allowing participants to concentrate fully on the haptic stimuli without external interference. This setup ensured that the collected data reflected the system’s performance under consistent conditions and enabled reliable assessment of tactile pattern recognition.

\begin{figure}[h]
\centerline{\includegraphics[width=0.5\textwidth]{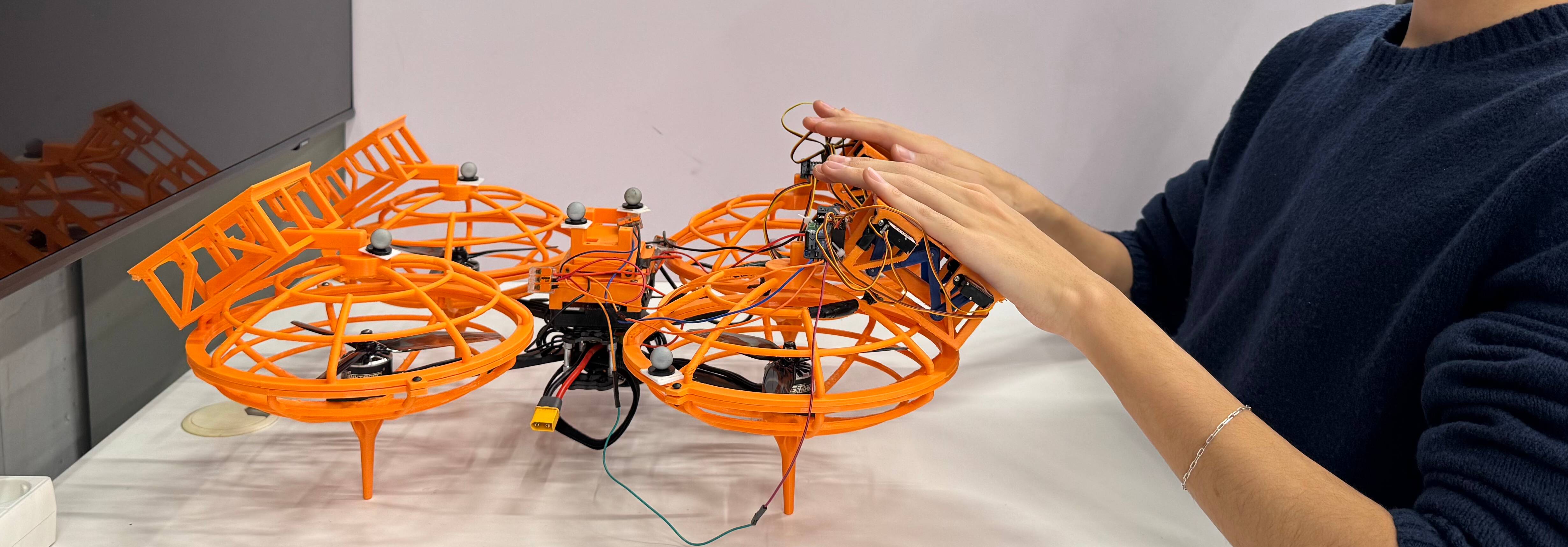}}
\caption{Tactile Pattern Experimental Setup}
\label{fig:ground_experiment}
\vspace{-0.4cm}
\end{figure}

A total of eight individuals participated in the study (seven men and one woman), with ages ranging from 22 to 35 years (mean age: 27.2 ± 3.7 years). All participants were right-handed to ensure consistency in interaction and feedback localization.

Prior to the experiment, each participant signed an informed consent form and received a comprehensive briefing outlining the study's goals, procedures, and interaction flow. This ensured a clear understanding of the task and mitigated the risk of confusion during the experimental trials.

\subsection{Experimental Procedure}

Before the testing phase, participants underwent a structured training session designed to familiarize them with the FlyHaptics system and the five tactile patterns. During this session, the experimenter explained how the system operates and demonstrated each tactile pattern.

To guarantee consistent feedback, the FlyHaptics device was calibrated before training by aligning the five-bar linkage mechanisms for accurate delivery of static patterns. Participants then placed their hands on the interface, where each pattern was delivered three times in sequence. A printed reference sheet showing all five tactile configurations was provided throughout the training phase, allowing participants to visually match the sensations they experienced to the corresponding labeled patterns.

During the main experiment, each participant remained seated at a desk and placed both hands on the haptic interface. The FlyHaptics system randomly presented each of the five predefined tactile patterns five times, ensuring a total of 25 trials per participant with an unbiased distribution of pattern order.

After each stimulus presentation, participants verbally reported the pattern they perceived. This method allowed for immediate data collection while maintaining focus on the tactile experience and minimizing distractions that written responses might introduce. The experimental configuration, including participant hand placement and system interaction, is shown in Fig. \ref{fig:ground_experiment}.

Following the trials, all participants took part in a debriefing session to share their feedback on the system. They were asked to describe any difficulties in recognizing specific patterns, note similarities between sensations, and provide general impressions regarding usability and comfort.

The debriefing followed a structured interview format to ensure thorough qualitative data collection. This post-experiment step provided valuable insight not only into recognition accuracy but also into user perception of the system’s intuitiveness, physical ergonomics, and overall haptic quality.

\subsection{Flight Test}

Following the completion of the user study, a dedicated flight test was conducted to evaluate how drone-induced vibrations and dynamic system behavior affected the haptic feedback experience. This test was carried out in a controlled indoor environment equipped with a Vicon-based localization system, which provided high-precision position tracking and supported stable flight control during the experiment.

The primary goal of the flight test was to assess whether airborne operation influenced the reliability and clarity of haptic sensations. Specifically, the test examined the presence of unwanted mechanical vibrations, positional drift, or oscillations that could interfere with the precision and fidelity of the tactile output.

By operating the system under real flight conditions, the test validated the practicality of FlyHaptics for mid-air applications. Results confirmed that the haptic feedback remained consistent, accurate, and safe during stable hover, demonstrating that FlyHaptics is capable of delivering controlled tactile interactions while in flight—thus reinforcing its potential for integration into immersive and mobile haptic environments

\section{Results}

To evaluate FlyHaptics’ effectiveness in delivering clear and distinguishable haptic feedback, participant responses were analyzed using a confusion matrix (see Table \ref{table:confusion_matrix}). The matrix provides a visual breakdown of how often each tactile pattern was correctly identified and where misclassifications occurred. This enabled a detailed assessment of recognition accuracy and helped reveal whether certain patterns were more easily distinguishable than others.

To examine performance differences across the five predefined tactile patterns, a single-factor repeated-measures ANOVA was conducted. This test was selected to determine whether recognition rates varied significantly across conditions while accounting for repeated responses from the same participants. By controlling for individual variability, the analysis ensured that the observed differences—if any—reflected true effects of the pattern characteristics rather than participant-specific biases.

Statistical computations were performed using the Pingouin and Statsmodels libraries, both widely adopted for behavioral research and human-computer interaction studies. A significance threshold of $\alpha < 0.05$ was used to identify statistically meaningful differences.

\begin{table}[]

\caption{Confusion Matrix for Actual and Perceived Tactile Patterns for All Subjects.}
\label{table:confusion_matrix}
\centering{
\begin{tabular}{|cc|ccccc|}
\hline
\multicolumn{2}{|c|}{}                               & \multicolumn{5}{c|}{Answers   (Predicted Class)}                                                                                                                                                                                                                                                                                                                \\ \cline{3-7} 
\multicolumn{2}{|c|}{\multirow{-2}{*}{\%}}           & \multicolumn{1}{c|}{1}                                                   & \multicolumn{1}{c|}{2}                                                   & \multicolumn{1}{c|}{3}                                                   & \multicolumn{1}{c|}{4}                                                   & 5                                                   \\ \hline
\multicolumn{1}{|c|}{}                           & 1 & \multicolumn{1}{c|}{\cellcolor[HTML]{506B76}{\color[HTML]{FFFFFF} 0.70}} & \multicolumn{1}{c|}{\cellcolor[HTML]{FFFFFF}-}                        & \multicolumn{1}{c|}{\cellcolor[HTML]{F3F5F6}0.05}                        & \multicolumn{1}{c|}{\cellcolor[HTML]{E6EAEC}0.10}                        & \cellcolor[HTML]{DAE0E2}0.15                        \\ \cline{2-7} 
\multicolumn{1}{|c|}{}                           & 2 & \multicolumn{1}{c|}{\cellcolor[HTML]{F3F5F6}0.05}                        & \multicolumn{1}{c|}{\cellcolor[HTML]{254654}{\color[HTML]{FFFFFF} 0.88}} & \multicolumn{1}{c|}{\cellcolor[HTML]{F3F5F6}0.05}                        & \multicolumn{1}{c|}{\cellcolor[HTML]{F9FAFB}{\color[HTML]{333333} 0.03}} & \cellcolor[HTML]{FFFFFF}-                        \\ \cline{2-7} 
\multicolumn{1}{|c|}{}                           & 3 & \multicolumn{1}{c|}{\cellcolor[HTML]{F3F5F6}0.05}                        & \multicolumn{1}{c|}{\cellcolor[HTML]{F9FAFB}0.03}                        & \multicolumn{1}{c|}{\cellcolor[HTML]{1E404F}{\color[HTML]{FFFFFF} 0.90}} & \multicolumn{1}{c|}{\cellcolor[HTML]{F9FAFB}0.03}                        & \cellcolor[HTML]{FFFFFF}-                        \\ \cline{2-7} 
\multicolumn{1}{|c|}{}                           & 4 & \multicolumn{1}{c|}{\cellcolor[HTML]{FFFFFF}-}                        & \multicolumn{1}{c|}{\cellcolor[HTML]{FFFFFF}-}                        & \multicolumn{1}{c|}{\cellcolor[HTML]{F9FAFB}0.03}                        & \multicolumn{1}{c|}{\cellcolor[HTML]{0B3040}{\color[HTML]{FFFFFF} 0.98}} & \cellcolor[HTML]{FFFFFF}-                        \\ \cline{2-7} 
\multicolumn{1}{|c|}{\multirow{-5}{*}{Patterns}} & 5 & \multicolumn{1}{c|}{\cellcolor[HTML]{F9FAFB}{\color[HTML]{333333} 0.03}} & \multicolumn{1}{c|}{\cellcolor[HTML]{F9FAFB}0.03}                        & \multicolumn{1}{c|}{\cellcolor[HTML]{F9FAFB}{\color[HTML]{333333} 0.03}} & \multicolumn{1}{c|}{\cellcolor[HTML]{F3F5F6}0.05}                        & \cellcolor[HTML]{254654}{\color[HTML]{FFFFFF} 0.88} \\ \hline
\end{tabular}}
\end{table}

The ANOVA yielded $F(4, 35) = 1.4665$ with a $p$-value of $0.2332$, indicating that there were no statistically significant differences in recognition accuracy across the five tactile patterns. This result suggests that each pattern was perceived with similar reliability, and no individual pattern was significantly more or less recognizable than the others.

The average recognition accuracy across all patterns was 86.5

The absence of significant differences between patterns reinforces the uniformity and balance of the haptic rendering mechanism. All five patterns were delivered with comparable clarity, and none stood out as particularly confusing or difficult to recognize. This result underscores the system's robustness and suitability for applications requiring consistent and repeatable haptic interactions.

While the overall accuracy was high, the confusion matrix reveals some minor errors, pointing to occasional misclassifications between similar patterns. These instances suggest that subtle adjustments to the linkage configurations or vibration parameters may further enhance pattern distinctiveness.

In addition, expanding the current static pattern evaluation to include dynamic haptic feedback would offer deeper insights into FlyHaptics’ capabilities in more complex interaction scenarios. Introducing time-varying stimuli could simulate real-world tactile conditions more effectively, broadening the system’s applicability to domains such as virtual reality, teleoperation, and remote robotic manipulation.


\end{document}